\documentclass[prl,showpacs,preprintnumbers,twocolumn]{revtex4}
\usepackage{graphicx}
\begin{document}
\bibliographystyle{apsrev}
\title{Magnetic inversion symmetry breaking and ferroelectricity in ${\bf TbMnO_{3}}$}

\author{M. Kenzelmann$^{1,2,3}$, A.~B. Harris$^{4}$, S. Jonas$^{3}$,
C. Broholm$^{3,5}$, J. Schefer$^{2}$,  S.~B. Kim$^{6}$, C.~L.
Zhang$^{6}$, S.-W. Cheong$^{6}$, O.~P. Vajk$^{5}$ and J.~W.
Lynn$^{5}$}

\affiliation{(1) Laboratory for Solid State Physics, ETH
H\"{o}nggerberg, CH-8093 Z\"{u}rich, Switzerland\\(2) Laboratory for
Neutron Scattering, ETH Z\"{u}rich \& Paul Scherrer Institute,
CH-5232 Villigen, Switzerland\\(3) Department of Physics and
Astronomy, Johns Hopkins University, Baltimore, Maryland 21218\\(4)
Department of Physics and Astronomy, University of Pennsylvania,
Philadelphia, PA 19104S\\(5) NIST Center for Neutron Research,
National Institute of Standards and Technology, Gaithersburg,
Maryland 20899\\(6) Department of Physics \& Astronomy, Rutgers
University, 136 Frelinghuysen Rd., Piscataway, NJ 08854}

\date{\today}

\begin{abstract}
${\rm TbMnO_3}$ is an orthorhombic insulator where incommensurate
spin order for temperature $T_N < 41\;\mathrm{K}$ is accompanied by
ferroelectric order for $T < 28\;\mathrm{K}$. To understand this, we
establish the magnetic structure above and below the ferroelectric
transition using neutron diffraction. In the paraelectric phase, the
spin structure is incommensurate and longitudinally-modulated. In
the ferroelectric phase, however, there is a transverse
incommensurate spiral. We show that the spiral breaks spatial
inversion symmetry and can account for magnetoelectricity in ${\rm
TbMnO_3}$.
\end{abstract}
 \maketitle

The coexistence of antiferromagnetism and ferroelectricity in solid
materials is rare, and much rarer still is a strong coupling between
these two order parameters \cite{Smolenskii,Schmid}. In non-magnetic
perovskites like ${\rm BaTiO_3}$, ferroelectricity is driven by a
hybridization of empty d orbitals with occupied p orbitals of the
octahedrally coordinated oxygen ions \cite{Cohen}. This mechanism
requires empty d orbitals and thus cannot lead to magnetic
ferroelectric materials. In materials such as ${\rm BiMnO_3}$, lone
$s^2$ electron pairs can lower their energy by hybridizing with
empty p orbitals \cite{Atanasov}. While it leads to coexistence of
magnetic order with electric polarization at low temperatures, the
very different ordering temperatures show that the two order
parameter lower the symmetry of the systems through distinctly
different collective effects.\par

${\rm TbMnO_3}$ is an antiferromagnet which contains ${\rm Mn^{3+}}$
ions with occupied d orbitals and no lone s$^2$ cation
\cite{Kimura}. So, as for a number of recently discovered
multiferroics \cite{Lottermoser,Hur,Kobayashi}, neither of the
mechanisms described above can explain the coexistence of magnetic
and electric order. In these materials, a magnetic field of a few
Tesla can switch the direction of the electric polarization
\cite{Kimura} - proof of a strong direct coupling between the
magnetic and electric polarization. Rare earth manganese oxides show
a plethora of exciting phenomena which arise from competing
interactions. ${\rm LaMnO_3}$ is the parent compound to a series of
materials featuring colossal magnetoresistance, and shows orbital
ordering of its $e_g$ orbital, giving rise to layered
antiferromagnetic ordering. With decreasing rare-earth size, there
is a tendency towards incommensurate magnetic order. Ferroelectric
ground states were recently predicted for doped manganite of the
type ${\rm R_{1-x}Ca_xMnO_3}$, but they have not yet been observed
\cite{Efremov}. The coexistence and strong coupling of
ferroelectricity and antiferromagnetism in ${\rm TbMnO_3}$ suggests
the presence of a non-conventional coupling mechanism involving
competing spin interactions, and it is thus of both practical and
fundamental importance.\par

To develop a microscopic theory of the new coupling mechanism,
unambiguous determination of the symmetry of the magnetic order is
essential. In this Letter, we present neutron diffraction
measurements of orthorhombic ${\rm TbMnO_3}$, which determine the
magnetic ground states and the phase diagram as a function of
temperature and a magnetic field ${\bf H} || {\bf a}$. We find that
the ferroelectric phase transition coincides with a magnetic
transition from a longitudinal incommensurate structure to an
incommensurate spiral structure that breaks spatial inversion
symmetry. We show that a recent theory proposed for axial non-axial
parity breaking \cite{Harris,LawesRapid} predicts the observed
orientation of the ferroelectric polarization based on the symmetry
of the magnetic structure that we report.\par

${\rm TbMnO_{3}}$ crystals were grown using an optical floating zone
furnace. A measurement of the temperature dependence of the
dielectric constant of the two samples which we studied confirmed
ferroelectricity below $T$=$26\;\mathrm{K}$ and $T$=$28\;\mathrm{K}$
for sample 1 and 2, respectively. The discrepancy may result from
slightly differing oxygen partial pressure during annealing. The
space group is \#62, and in the Pbnm setting the lattice parameters
are $a$=$5.3\AA$, $b$=$5.86\AA$ and $c$=$7.49\AA$. The measurements
were performed on two single-crystals weighing $40\;\mathrm{mg}$ and
$220\;\mathrm{mg}$, respectively, with the BT2 and SPINS
spectrometers at NIST, and the TriCS 4-circle diffractometer at
PSI.\par

\begin{figure}
\begin{center}
  \includegraphics[height=6cm,bbllx=60,bblly=230,bburx=495,
  bbury=550,angle=0,clip=]{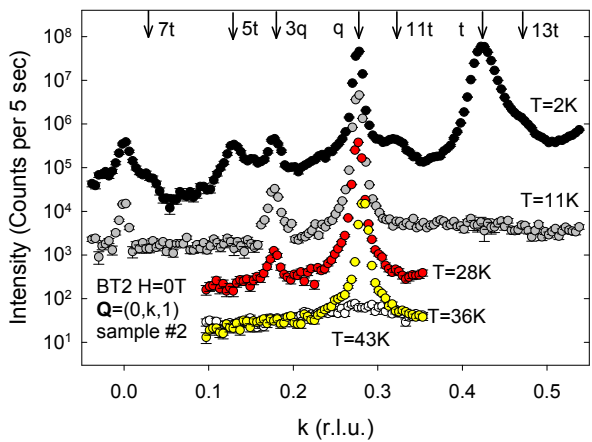}
  \caption{Diffraction intensity along the $(0,k,1)$ direction
  for different temperatures (with a factor 10 offset between
  the sub-$T_N$ data sets), showing fundamental and
  higher order Fourier components of the magnetic ordering.
  The $3t$ and $9t$-order peaks associated with the Tb
  order occur at $k=0.725$ and $0.175$, respectively, and are
  difficult to identify unambiguously due to their proximity
  to strong peaks.}
  \vspace{-0.9cm}
  \label{kscansT}
\end{center}
\end{figure}

\begin{figure}
\begin{center}
  \includegraphics[height=6.6cm,bbllx=45,bblly=245,bburx=470,
  bbury=640,angle=0,clip=]{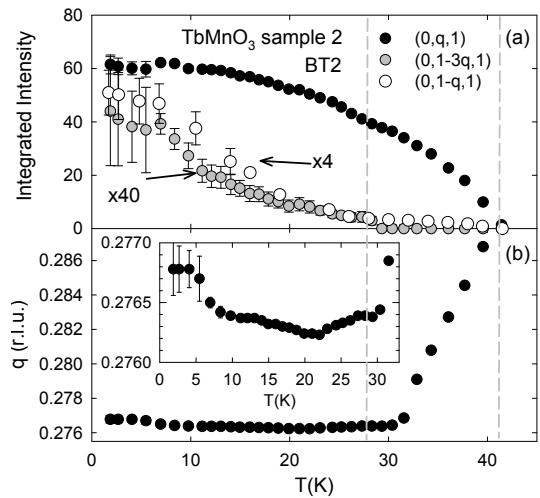}
  \caption{(a) T-dependence of the integrated intensity
  of the $(0,q,1)$ Bragg peak and the associated third order peak
  at $(0,1-3q,1)$. (b) T-dependence of the wave-vector
  $q$, corrected for the T-dependence of the b lattice
  parameter, showing the absence of a lock-in transition at
  $T$=$28\;\mathrm{K}$.}
  \vspace{-0.8cm}
  \label{Fig2IntqvsT}
\end{center}
\end{figure}

The wave-vector dependence of the diffraction intensity along the
$(0,k,1)$ direction of reciprocal space (Figs.~\ref{kscansT} and
\ref{Fig2IntqvsT}) illustrates the temperature dependence of
magnetic order in ${\rm TbMnO_{3}}$. ${\rm Mn^{3+}}$ spins develop
long-range order at $T_N$=$41\;\mathrm{K}$ \cite{Quezel,Kajimoto}.
Immediately below the transition temperature, magnetic Bragg peaks
are observed at $(0,1-q,1)$ and $(0,q,1)$ positions. At
$35\;\mathrm{K}$ the magnetic order is described by a single Fourier
component associated with a wave-vector $(0,q,0)$ with $q=0.27$, and
no apparent higher order Fourier components. Upon cooling below
$T$=$28\;\mathrm{K}$, third-order magnetic Bragg peaks $(0,1-3q,0)$
appear (Fig.~\ref{kscansT}), indicating step-like modulation of the
magnetic moment. The inset to Fig.~\ref{Fig2IntqvsT}b shows that the
wave vector continues to vary with temperature throughout the
ferroelectric phase.\par

Below $T$=$7\;\mathrm{K}$, additional magnetic peaks appear that are
associated with Tb moments and indicate that their interactions
favor a different ordering wave-vector than for the Mn moments. At
$T$=$2\;\mathrm{K}$, there is a strong peak at $(0,t,1)$ with
$t=0.425$, and several higher-order peaks as indicated in
Fig.~\ref{kscansT}. The many strong odd high-order reflections
indicate that the order associated with the $(0,t,0)$ wave-vector is
strongly distorted (bunched structure) as expected for anisotropic
rare-earth magnets. The correlation length and the incommensuration
of the Tb order at low-temperatures is sample dependent, as is often
found in systems with phase transitions between incommensurate
structures. The correlation length along the b-axis was $58(20)$ and
$280(20)\AA$, and $t$ was $0.41$ and $0.425$ in sample $1$ and $2$,
respectively.\par

\begin{figure}
\begin{center}
  \includegraphics[height=7.2cm,bbllx=86,bblly=257,bburx=480,
  bbury=640,angle=0,clip=]{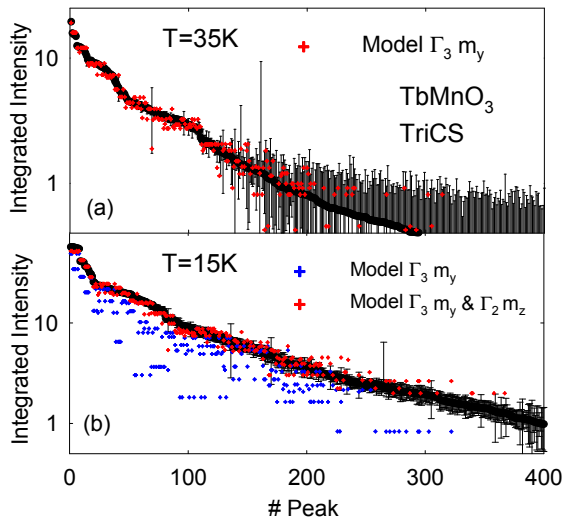}
  \caption{Integrated intensities of magnetic Bragg peaks
  measured at $T$=$15\;\mathrm{K}$ and $T$=$35\;\mathrm{K}$
  using TriCS are compared to various magnetic structures. Peaks
  are sorted by decreasing measured intensity.
  (a) The $T$=$35\;\mathrm{K}$ structure is described by a single
  irreducible representation, $\Gamma_3$, and the moments point
  along the b-axis. (b) The $T$=$15\;\mathrm{K}$ magnetic
  structure is a spiral described by a $y$-component of $\Gamma_3$
  and an $x$-component of $\Gamma_2$ with a $\pi/2$ phase shift.}
  \vspace{-0.9cm}
  \label{structurefits}
\end{center}
\end{figure}

\begin{figure}
\begin{center}
  \includegraphics[height=7cm,bbllx=15,bblly=0,bburx=494,
  bbury=420,angle=0,clip=0]{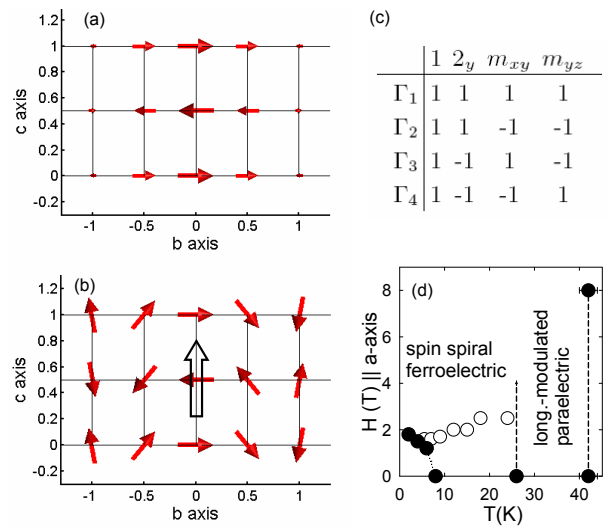}
  \caption{Schematic of the magnetic structure at (a)
  $T$=$35\;\mathrm{K}$ and (b) $T$=$15\;\mathrm{K}$, projected
  onto the b-c plane. Filled arrows indicate direction and
  magnitude of Mn moments. The longitudinally-modulated phase
  (a) respects inversion symmetry along the c axis, but the
  spiral phase (b) violates it, allowing an electric polarization
  (unfilled arrow). (c) Irreducible representation of the group
  $G_{\bf k}$ for the incommensurate magnetic structure with
  ${\bf k}=(0,q,0)$. (d) Phase diagram as a function of
  temperature and field applied along the a-axis. Full circles
  indicate second order phase transitions. Open circles indicate
  the characteristic field for reduction of magnetic
  $(0,1-q,1)$ Bragg scattering from Tb moments by $50\%$
from its zero-field intensity.}
  \vspace{-0.8cm}
  \label{MagnStructure}
\end{center}
\end{figure}

The magnetic structures at $T$=$15\;\mathrm{K}$, and
$35\;\mathrm{K}$ were determined from up to 922 first-order magnetic
Bragg peaks each. Only first order peaks were included in the
refinement as these are sufficient to determine the symmetry of the
magnetic structure. Representational analysis was used to find the
irreducible representations that describe the magnetic structures.
The structure at $T$=$35\;\mathrm{K}$ in the high-temperature
incommensurate (HTI) phase can be described by a single irreducible
representation. The best fit with $\chi^2=0.86$ was obtained for a
structure described by representation $\Gamma_3$, and the excellent
agreement between the model and the data is shown in
Fig.~\ref{structurefits}a). The magnetic structure is described by
${\bf m}^{\rm Mn}_3 = \left(0.0(8),2.90(5),0.0(5)\right)\mu_B$ and
${\bf m}^{\rm Tb}_3 = \left(0,0,0.0(4)\right)\mu_B$ where the
subscript denotes the irreducible representation indicated in
Fig.~\ref{MagnStructure}c. The magnetic structure is
longitudinally-modulated with moment along the b-axis, as
illustrated in Fig.~\ref{MagnStructure}a and consistent with an
earlier study \cite{Quezel}. The absence of observable higher order
peaks indicates that the magnetic structure at $T$=$35\;\mathrm{K}$
is sinusoidally modulated.\par

Two irreducible representations are required to describe the
magnetic structure at $T$=$15\;\mathrm{K}$ in the low-temperature
incommensurate (LTI) phase. We found best agreement with
$\chi^2=2.19$ for magnetic ordering involving $\Gamma_2$ and
$\Gamma_3$, as shown in Fig.~\ref{structurefits}b. Fits using the
$\Gamma_1$ and $\Gamma_3$, or the $\Gamma_3$ and $\Gamma_4$
representation pairs led to $\chi^2=14.5$ and higher, and can thus
be excluded. Neglecting higher order reflections, the magnetic
structure is given by ${\bf m}^{\rm
Mn}_3=\left(0.0(5),3.9(1),0.0(7)\right)\mu_B$, ${\bf m}^{\rm Tb}_3 =
\left(0,0,0(1)\right)\mu_B$, ${\bf m}^{\rm Mn}_2
=\left(0.0(1),0.0(8),2.8(1)\right)\mu_B$ and ${\bf m}^{\rm Tb}_2  =
\left(1.2(1),0(1),0\right)\mu_B$. The experiment was not sensitive
to the phase between the $y$ and $z$-component of the Mn moment.
From the size of the moment, however, we deduce that the Mn moments
form an elliptical spiral. The data did not favor a phase difference
between the Tb and Mn moments, so these phases remain undetermined.
Symmetry splits the Tb moments into two orbits which representation
theory normally treats as independent. However, as suggested by
Landau theory \cite{Harris}, we took these two Tb amplitudes to be
identical. The phase between the two orbits was found to be to
$1.3(3)\pi$. The greatly improved fit is evidence that the Tb
sublattice carries significant magnetization in the LTI phase,
presumably as a consequence of the exchange field from the ordered
Mn sublattice.\par

\begin{figure}
\begin{center}
  \includegraphics[height=5.5cm,bbllx=38,bblly=218,bburx=545,
  bbury=560,angle=0,clip=]{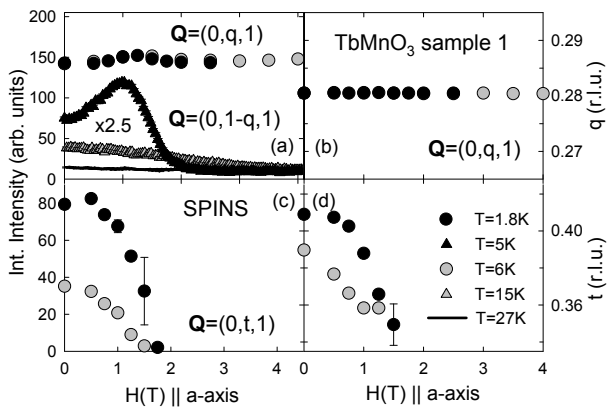}
  \caption{Field dependence of magnetic Bragg scattering
  from ${\rm TbMnO_3}$. (a) and (b) show data for the
  incommensurate peaks that occur for $T<41\;\mathrm{K}$.
  (a) The $(0,q,1)$ peak that is mostly
  sensitive to staggered magnetization on Mn sites and
  the $(0,1-q,1)$ peak that is sensitive to staggered
  magnetization on Tb sites. (c) and (d) show
  data for the incommensurate peaks that develop for
  $T<7\;\mathrm{K}$.}
  \vspace{-0.8cm}
  \label{Fig5fieldscans}
\end{center}
\end{figure}

Fig.~\ref{Fig5fieldscans}(a-b) shows the field dependence of the
$(0,q,1)$ magnetic Bragg reflection, which arises from the Mn spin
spiral. Both the position and the intensity are field independent to
within errorbar - evidence that the structure remains a spiral up to
at least $6\;\mathrm{T}$. Our calculations show that the intensity
should drop by $7\%$ if the $z$-component of $\Gamma_2$ were
extinguished. In contrast, no decrease is observed to within an
error bar of $2\%$ between $0$ and $6\;\mathrm{T}$.\par

The $(0,1-q,1)$ Bragg reflection shown in Fig.~\ref{Fig5fieldscans}a
arises from Mn $\Gamma_3$ magnetization along the a-axis and from Tb
$\Gamma_2$ magnetization along the a-axis. Because the $x$-component
of the Mn moment is small, the $(0,1-q,1)$ Bragg reflection is
particularly sensitive to Tb order. For $T<28\;\mathrm{K}$ the
$(0,1-q,1)$ intensity is suppressed by a field ${\bf H}||{\bf a}$
confirming that the modulated Tb moment is oriented along that
direction. Below the Tb ordering temperature, the field dependent
magnetic Bragg intensity has a finite-field maximum
(Fig.~\ref{Fig5fieldscans}a), indicating a spin-flop transition.\par

We collected $51$ magnetic Bragg peaks at $T$=$4\;\mathrm{K}$ and
$H$=$4\;\mathrm{T}$ along ${\bf a}$ to determine the magnetic
structure at low temperatures above the critical field for $(0,t,0)$
Tb order (Fig.~\ref{Fig5fieldscans}c). The magnetic structure can be
described by $\Gamma_2$ and $\Gamma_3$ with $\chi^2= 3.81$ or by
$\Gamma_1$ and $\Gamma_3$ with $\chi^2= 4.19$. Since a field along
the a-direction disfavors antiparallel spin alignment in the same
direction as in the $\Gamma_1$-$\Gamma_3$ structure, we infer that
the structure is given by ${\bf m}^{\rm Mn}_3 =
\left(0.3(4),4.7(3),0.0(5)\right)\mu_B$, ${\bf m}^{\rm Tb}_3  =
\left(0,0,0.0(3)\right)\mu_B$, ${\bf m}^{\rm Mn}_2 =
\left(0.0(2),0.0(4),3.0(3)\right)\mu_B$ and ${\bf m}^{\rm Tb}_2  =
\left(0.3(2),0.0(4),0\right)\mu_B$. This result suggests that the
spin spiral structure is more stable for fields along the a-axis
than for fields along the b-axis \cite{Kimura}.\par

Harris {\it et al.} \cite{Harris,LawesRapid} recently showed that
insulators with axial-non-axial parity breaking magnetic phase
transitions must also be electrically polarized. Given the magnetic
structure and the temperature dependence of the magnetic order
parameter, the theory predicts the direction and temperature
dependence of the electric polarization resulting from a symmetry
allowed trilinear coupling term. In the following we show that this
theory correctly accounts for the direction of the electric
polarization in the LTI phase of ${\rm TbMnO_3}$, and the absence of
electric polarization in the HTI phase. The trilinear
magnetoelectric coupling term in the Landau free energy expansion is
written as $V=\sum_{\rm uv\gamma} a_{\rm uv\gamma}\sigma_{\rm u}(k)
\sigma_{\rm v}(-k) P_{\rm \gamma}$. Here $\sigma_{\rm u}(k)$ is the
magnetic order parameter of irreducible representation $\Gamma_u$,
$P_{\rm \gamma}$ is the electric polarization along the the $\gamma$
crystallographic direction and $a_{\rm uv\gamma}$ parametrizes the
strength of the interaction between the electric and magnetic order
parameters. In the HTI phase, the magnetic order is described by
only one irreducible representation, $\Gamma_3$ and due to the high
symmetry of the Mn moments, it is possible to define $\sigma_u$ such
that under inversion $\sigma_{\rm u}(k) \rightarrow \sigma_{\rm
u}(k)^{\star}$. The trilinear coupling thus consists only of terms
such as $V = \sum_{\rm \gamma} a_{\rm \gamma} |\sigma_{\rm 3}(k)|^2
P_{\rm \gamma}$. Since interactions in a Landau expansion must have
the symmetry of the paramagnetic phase, this interaction must be
invariant under inversion. This requires that $a_{\rm \gamma}$
vanishes, so that $P_{\rm \gamma}=0$ in the HTI phase. The
conclusion remains valid for Tb order with no restriction on the
phase between its two orbits.\par

For the LTI phase, which is described by $\Gamma_2$ and $\Gamma_3$,
there are, however, additional terms such as $V=\sum_{\rm \gamma}
b_{\rm \gamma} \sigma_{\rm 2}(k) \sigma_{\rm 3}(-k) P_{\rm \gamma} +
c. c.$, where c. c. denotes the complex conjugate. For $V$ to be an
invariant, $P_{\rm \gamma}$ must transform as the product of
$\Gamma_2$ and $\Gamma_3$. That is, the electric polarization must
be even under $1$ and $m_{\rm yz}$, and odd under $2_{\rm y}$ and
$m_{\rm xy}$. This condition can only be satisfied for an electric
polarization along the c-axis. Previous dielectric experiments have
shown that the electric polarization that develops below the
transition to the LTI phase is indeed oriented along the c-axis.\par

The coupling of the magnetic order parameter to electric
polarization in ${\rm TbMnO_3}$  \cite{Kimura} is similar to that in
${\rm Ni_3V_2O_8}$, which adopts two different incommensurate
magnetic structures \cite{LawesPRL}, one of them ferroelectric and
described by two irreducible representations. The correct prediction
of the electric polarization for both ${\rm TbMnO_3}$ and ${\rm
Ni_3V_2O_8}$ (Ref.~\onlinecite{LawesRapid}~) suggests that
magnetoelectricity resulting from a trilinear magnetoelectric
coupling term may be commonplace in insulating transition metal
oxides with non-collinear incommensurate structures. Accordingly we
find less compelling the suggestion \cite{Kimura} that the
appearance of ferroelectricity is associated with an incommensurate
to commensurate phase transition. Indeed,  there is no evidence of a
lock-in transition (see Fig.~\ref{Fig2IntqvsT}b) and it therefor
appears to be irrelevant from the point of view of ferroelectricity
whether the modulated magnetic order is commensurate or truly
incommensurate.\par

In summary, we have determined the magnetic structure of the
paraelectric and ferroelectric phases of ${\rm TbMnO_3}$. We showed
that the paraelectric, magnetically incommensurate phase has
sinusoidally-modulated collinear magnetic order that does not break
inversion symmetry. The ferroelectric phase, however, has
non-collinear incommensurate magnetic order described by two
irreducible representations, which explicitly breaks inversion
symmetry and thus gives rise to electric polarization. The
qualitative aspects of magnetoelectric effects in ${\rm TbMnO_3}$
appear to be accounted for by a trilinear coupling term in a Landau
free energy expansion as proposed by Harris {\it et al.}
\cite{Harris,LawesRapid}. Understanding the magnitude of the effect
will require experimental as well as theoretical work to track
lattice, charge, and orbital degrees of freedom through
axial-non-axial parity breaking phase transition in insulators such
as ${\rm TbMnO_3}$. Apart from the fundamental challenge, improved
understanding of magnetoelectricity in these systems may help to
produce materials for room-temperature applications.


\begin{acknowledgments}
We thank A. Aharony, O. Entin-Wohlman, and A. Ramirez for helpful
discussions. This work was supported by the Swiss National Science
Foundation under Contract No. PP002-102831. Work at Johns Hopkins
University was supported by the DoE through DE-FG02-02ER45983. Work
at University of Pennsylvania was supported by the U.S.-Israel
Binational Science Foundation under Grant number 2000073. Work at
Rutgers University was supported by the NSF-DMR-MRSEC-00-080008.
This work is based on experiments performed at the Swiss spallation
neutron source SINQ, Paul Scherrer Institute, Villigen, Switzerland.
The work at SPINS is based upon activities supported by the National
Science Foundation under Agreement No. DMR-9986442.
\end{acknowledgments}


\end{document}